\documentclass[aps,prb,twocolumn,showpacs,preprintnumbers,amsmath,amssymb,superscriptaddress]{revtex4}

\usepackage{graphicx}
\usepackage{dcolumn}
\usepackage{bm}

\begin{document}


\title{Magnetic anisotropy of epitaxial (Ga,Mn)As on (113)A GaAs}

\author{Wiktor Stefanowicz}
\affiliation{Institute of Physics, Polish Academy of Science, al.~Lotnik\'ow 32/46, PL-02-668 Warszawa, Poland}
\affiliation{Laboratory of Magnetism, University of Bia{\l}ystok, ul.~Lipowa 41, PL-15-424 Bia{\l}ystok, Poland}
\author{Cezary \'Sliwa}
\affiliation{Institute of Physics, Polish Academy of Science, al.~Lotnik\'ow 32/46, PL-02-668 Warszawa, Poland}
\author{Pavlo Aleshkevych}
\affiliation{Institute of Physics, Polish Academy of Science,
al.~Lotnik\'ow 32/46, PL-02-668 Warszawa, Poland}
\author{Tomasz Dietl}
\affiliation{Institute of Physics, Polish Academy of Science, al.~Lotnik\'ow 32/46, PL-02-668 Warszawa, Poland}
\affiliation{Institute of Theoretical Physics, University of Warsaw, PL-00-681 Warszawa, Poland}
\author{Matthias D\"oppe}
\affiliation{Department of Physics, University Regensburg, 93040 Regensburg, Germany}
\author{Ursula Wurstbauer}
\affiliation{Department of Physics, University Regensburg, 93040 Regensburg, Germany}
\author{Werner Wegscheider}
\affiliation{Department of Physics, University Regensburg, 93040 Regensburg, Germany}
\author{Dieter Weiss}
\affiliation{Department of Physics, University Regensburg, 93040 Regensburg, Germany}
\author{Maciej Sawicki}
\affiliation{Institute of Physics, Polish Academy of Science, al.~Lotnik\'ow 32/46, PL-02-668 Warszawa, Poland}

\date{\today}

\begin{abstract}
The temperature dependence of magnetic anisotropy in
$(113)\mathrm{A}$ (Ga,Mn)As layers grown by molecular beam epitaxy
is studied by means of superconducting quantum interference device
(SQUID) magnetometry as well as by ferromagnetic resonance (FMR)
and magnetooptical effects. Experimental results are described
considering cubic and two kinds of uniaxial magnetic anisotropy.
The magnitude of cubic and uniaxial anisotropy constants is found
to be proportional to the fourth and second power of saturation
magnetization, respectively. Similarly to the case of (001)
samples, the spin reorientation transition from uniaxial
anisotropy with the easy axis along the $[\overline110]$ direction at
high temperatures to the biaxial $\langle 100\rangle$ anisotropy
at low temperatures is observed around 25~K. The determined values
of the anisotropy constants  have been confirmed by FMR studies.
As evidenced by investigations of the polar magnetooptical Kerr
effect, the particular combination of magnetic anisotropies allows
the out-of-plane component of magnetization to be
reversed by an in-plane magnetic field. Theoretical calculations
within the $p-d$ Zener model explain the magnitude of the
out-of-plane uniaxial anisotropy constant caused by epitaxial
strain  but do not explain satisfactorily the cubic anisotropy
constant. At the same time the findings point to the presence of
an additional uniaxial anisotropy of unknown origin. Similarly to
the case of (001) films, this additional anisotropy can be
explained by assuming the existence of a shear strain. However, in
contrast to the (001) samples, this additional strain has an out
of the (001) plane character.
\end{abstract}

\maketitle

\section{Introduction}
\label{sec: intro}

Since many decades, a lot of attention has been devoted to
ferromagnetic semiconductors.  More recently, the intense research
has been triggered by the synthesis of the (III,Mn)V diluted
magnetic semiconductor (Ga,Mn)As,\cite{Ohno:1996_APL} which has
become the canonical example of a dilute ferromagnetic
semiconductor.\cite{Matsukura:2002_B,Dietl:2008_B} It has been
demonstrated that a number of pertinent properties of this
material can be explained by the $p-d$ Zener
model.\cite{Dietl:2000_S,Dietl:2001_PRB,Jungwirth:2006_RMP,Dietl:2008_B}
Magnetic anisotropy of strained (Ga,Mn)As layers can be calculated
within this theory, and many experimental
studies\cite{Tang:2003_PRL,Welp:2003_PRL,Sawicki:2004_PRB,Sawicki:2005_PRB,Wang:2005_PRL,Liu:2006_g,Thevenard:2007_PRB,Gourdon:2007_PRB,Gould:2008_a}
were devoted to verify its predictions. However, despite these
intense studies, some important features of magnetic anisotropy in
this system are at present not completely understood.

An example of such a property is a rather strong in-plane uniaxial
magnetic anisotropy of epitaxial (Ga,Mn)As layers grown on GaAs
substrates of $(001)$ orientation. Owing to the presence of the
twofold symmetry axes $[100]$ and $[010]$, the in-plane
zinc-blende directions $[110]$ and $[\overline110]$ are expected
to be equivalent. Yet, as implied by the character of magnetic
anisotropy, the symmetry is lowered from $D_{2d}$ to $C_{2v}$,
possibly due to the growth-induced lack of symmetry between the
bottom and the top of the
layer,\cite{Welp:2004_APL,Sawicki:2004_PRB,Sawicki:2005_PRB} which
can be phenomenologically described by introducing a shear
strain.\cite{Sawicki:2005_PRB,Zemen:2009_PRB,Glunk:2009_PRB}

Since these symmetry considerations are limited to $(001)$ layers,
investigation of layers grown on substrates of other orientations
may not only allow to compare experimental observations with
predictions of the $p-d$ Zener model in a more general situation,
but also provide information from which conclusions on the nature
of the additional anisotropy can be drawn.

In this paper we present results of studies on
$\mathrm{Ga}_{1-x}\mathrm{Mn}_{x}\mathrm{As}$ layers grown by
low-temperature molecular beam epitaxy (MBE) on GaAs substrates
with the $(113)\mathrm{A}$ orientation.  Previously, magnetic
anisotropy in such films was probed at low temperatures by
magnetoresistance\cite{Wang:2005_PRB_b,Limmer:2006_PRB,Limmer:2006_MJ},
scanning Hall probe microscopy\cite{Pross:2006_JAP} and
ferromagnetic resonance
measurements.\cite{Limmer:2006_MJ,Bihler:2006_APL,Liu:2007_JPCM}
Experimental techniques employed here include superconducting
quantum interference device (SQUID) magnetometry, ferromagnetic
resonance (FMR), and polar magnetooptical Kerr effect (PMOKE). Our
measurements are carried out over a wide temperature and magnetic
field range. We find that magnetic anisotropy can be consistently
described taking into account three contributions: a uniaxial
anisotropy with the hard axis tilted from $[113]$ toward the
$[001]$ direction, an in-plane uniaxial anisotropy with the easy axis
along the $[\overline110]$ direction, and a cubic anisotropy with
easy $\langle 100 \rangle$ directions. The general form of
anisotropy is, therefore, similar to the case of (001) films, but
the direction of the hard axis is found to be neither along [001]
nor perpendicular to the film plane in the (113) case.  The
accumulated experimental results allow us to determine how the
three relevant magnetic anisotropy constants $K$ as well as the
tilt angle depend on the temperature. We find that the magnitudes
of energies corresponding to the competing cubic and uniaxial
anisotropies in the (001) plane depend, as could be expected, as
the fourth and second power of spontaneous magnetization $M(T)$,
respectively. In contrast, a complex dependence on $M(T)$ is
observed in the case of the energy characterizing the out-of-plane
uniaxial anisotropy. We assign this behavior to the
spin-splitting-induced and, hence, temperature dependent
redistribution of holes between the valence band subbands that are
characterized by different directions of the angular momentum and,
hence, of the easy axes.

\begin{figure}[bp]
   \begin{center}
        \includegraphics[width=0.95\columnwidth]{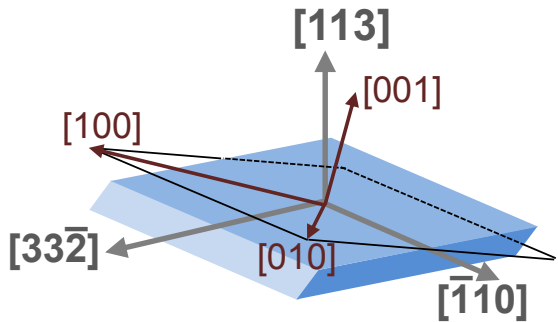}
   \end{center}
   \caption{(Color online) Crystallographic directions for a GaAs substrate of (113) orientation.}
   \label{fig: 1}
\end{figure}

In the theoretical part, we present a theory of magnetic
anisotropy in epitaxially strained layers of (Ga,Mn)As and related
systems within the $p-d$ Zener model. Our approach generalizes
earlier theories developed for (001)
films\cite{Dietl:2000_S,Dietl:2001_PRB,Abolfath:2001_PRB,Glunk:2009_PRB,Zemen:2009_PRB}
by allowing for an arbitrary crystallographic orientation of the
substrate. Similarly to previous
studies,\cite{Sawicki:2005_PRB,Zemen:2009_PRB} in order to explain
the experimental findings, we introduce an additional shear strain
whose three components constitute adjustable
parameters. We also take into account the Hamiltonian terms linear in $k$ and find
that they give a minor contribution to the magnitude of magnetic
anisotropy constants.

\section{Samples and experiment}
\label{seq: samples}

We study a 50~nm thick
$\mathrm{Ga}_{1-x}\mathrm{Mn}_{x}\mathrm{As}$ layer which has been
grown on a (113)A GaAs substrate (see Fig.~\ref{fig: 1}) by low
temperature MBE.\cite{Reinwald:2005_JCG} The total Mn
concentration of $x = 6.4\%$ has been determined by means of
secondary ion mass spectrometry, however only more than twice
lower value of an effective Mn concentration $x_{\text{eff}}$ can
be inferred from low temperature experimental saturation
magnetization, $M_{\text{exp}}$. This reduction of $x$ is
primarily caused by a presence of Mn interstitials. These point
defects act as double donors and form strongly coupled spin
singlet pairs with neighbor substitutional Mn
cations.\cite{Jungwirth:2006_RMP,Yu:2002_PRB,Blinowski:2003_PRB}
These pairs neither participate in the ferromagnetic order, nor
they contribute to $M$. Thus, the effective concentration of Mn
ions which generates $M_{\text{exp}}$ gets reduced to
$x_{\text{eff}} = x - 2x_{\text{I}}$, where $x_{\text{I}}N_0$ is
the concentration of the Mn interstitials and $N_0$ is the cation
concentration. However, the experimentally measured
$M_{\text{exp}}$ is further reduced by holes magnetization, $M_h$,
which is oppositely oriented to magnetization of Mn spins,
$M_{\text{Mn}}$, and so $M_{\text{Mn}}$ = $M_{\text{exp}}$  +
$|M_h|$ should be used to calculate $x_{\text{eff}}$, with $M_h$
being computed in the framework of the the mean-field $p-d$ Zener
model.\cite{Dietl:2000_S,Dietl:2001_PRB} We perform these
calculations in a self-consistent way taking the hole
concentration as $p = N_0(x- 3x_{\text{I}}) = N_0(3x_{\text{eff}}
- x)/2$, that is neglecting other charge compensating defects.

The open air post growth annealing at temperatures below or
comparable to the growth
temperature\cite{Hayashi:2001_APL,Edmonds:2002_APL} is a
frequently used procedure for improving material parameters of
(Ga,Mn)As, since the corresponding out-diffusion and passivation
of Mn interstitials\cite{Edmonds:2004_PRL} increases
$x_{\text{eff}}$, $p$, and eventually the Curie temperature
$T_{\mathrm{C}}$. Therefore in order to widen the parameter space
employed here to study the magnetic anisotropy in this compound we
investigate both the as-grown material (sample S1) and the samples
annealed at $200^{\circ}$C for 1.5 hour (sample S2) and 5 hours
(sample S3). Taking the determined values of $M_{\text{exp}}$ data
we end up with $x_{\text{eff}}$ = 2.7, 3.1, and 3.3\% and $p=
2.0$, 3.3, and $3.8 \times 10^{20}$~cm$^{-3}$ for which calculated
values of $T_{\mathrm{C}}$ = 46, 73, and 85~K compares favorably
with the experimentally established values of 65, 77, and 79~K,
for samples S1, S2, and S3, respectively.

Magnetic properties referred to above and described further on
have been obtained by employing a Quantum Design MPMS XL-5
magnetometer. A special demagnetization procedure has been
employed to minimize the influence of parasite fields on
zero-field measurements. The temperature dependence of remnant
magnetization, TRM, serves to obtain an overview of magnetic
anisotropy as well as to determine $T_{\mathrm{C}}$
(Sec.~\ref{sec: rem}). After cooling the sample across
$T_{\mathrm{C}}$ down to $5\,\mathrm{K}$ in the external magnetic
field of 0.1~T, the field is removed, allowing magnetization to
assume the direction along the closest easy axis. The magnitude of
the magnetization component along the magnet axis, TMR$_i$, is then
measured while heating, where $i$ indicates one of the three mutually
orthogonal directions of the magnetizing field, corresponding to
the surface normal $\bm{n}_1 = [113]$ and the two edges $\bm{n}_2
= [33\overline2]$, $\bm{n}_3 = [\overline110]$, as depicted in
Fig.~\ref{fig: 1}. Since, except to the immediate vicinity of the
spin reorientation transition, magnetization of (Ga,Mn)As films
tend to align in a single domain state, the measurements performed
for the three orthogonal axes provide the temperature dependence
of the magnetization magnitude and direction.

To study magnetic anisotropy in a greater detail, magnetic
hysteresis loops $M_i(H)$ have been recorded in external magnetic
field in the range of $\pm 0.5$~T along the three directions $i$.
The measurements have been carried out at various temperatures,
and the parameters of the anisotropy model (Sec.~\ref{sec:
phenomodel}) have been fitted to reproduce the magnetization data.
To cross check magnetic anisotropy constants obtained from SQUID
studies, FMR measurements have been performed at $\omega/2\pi =
9.3 \, \mathrm{GHz}$ and $T = 10$~K. We have performed angle
dependent measurements of the resonance field in the four
different crystallographic planes $(\overline110)$,
$(33\overline2)$, $(113)$ and $\frac{1}{2\sqrt{11}}(3-\sqrt{11},
3+\sqrt{11}, -2)$. As discussed in Sec.~\ref{sec: fmr}, the FMR
data are in a good agreement with the anisotropy model, employing
parameters determined from the SQUID measurements.

\section{Experimental results}
\subsection{Overview of magnetic anisotropy}
\label{sec: rem}

\begin{figure}[tp]
  \begin{center}
         \includegraphics[width=0.95\columnwidth]{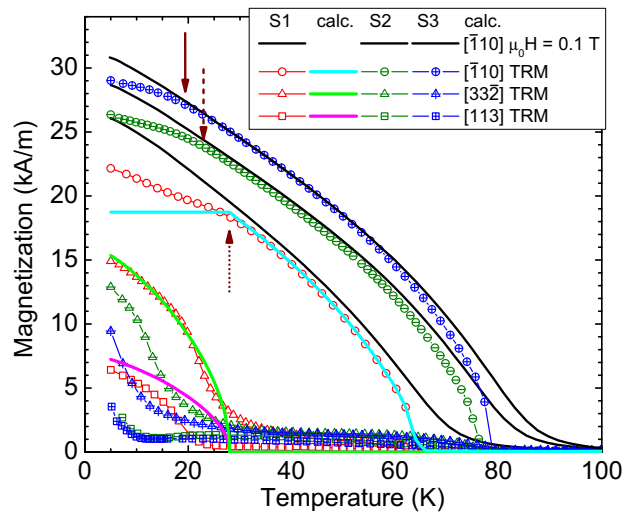}
  \end{center}
\caption{(Color online) Temperature dependence of remnant
magnetization components in all three samples (points). Black
lines: $M(H)$ at 0.1~T along $[\overline110]$ - the in-plane
uniaxial easy axis. Solid lines of lighter shades: the same
magnetization components calculated according to a model of only
two: uniaxial and biaxial magnetic anisotropies operating at (001)
plane and undergoing a spin reorientation transition at
temperatures $T^*$ (marked by arrows), as in (001) (Ga,Mn)As.}
  \label{fig:TRM}
\end{figure}

The TRM studies of all three samples are summarized in
Fig.~\ref{fig:TRM}. We immediately see that the
TRM$_{[\overline110]}$ component of TRM is the strongest for all
of the samples and that at elevated temperatures its magnitude is
nearly equal to the saturation magnetization $M(T)$, established
by the measurement in $\mu_0H=0.1$~T. Since the magnitude of the
other two magnetization components is vanishingly small, we find
that in this temperature range the in-plane uniaxial anisotropy
with the easy axis along $[\overline110]$ direction dominates.
This perfectly uniaxial behavior at $T \rightarrow T_{\mathrm{C}}$
allows us to use TRM$_{[\overline110]}$ to precisely determine
$T_{\mathrm{C}}$ in the studied samples (already given in the
previous section). On the other hand, below a certain temperature
$T^*$ (marked by arrow for every sample in Fig.~\ref{fig:TRM})
TRM$_{[\overline110]}$ gets visibly smaller than $M(T)$, and the
other in-plane TRM component, TRM$_{[33\overline2]}$, acquires
sizable values, followed at still lower temperatures by the
out-of-plane component TRM$_{[113]}$. This clearly indicates a
departure of the easy direction from the $[\overline110]$
direction below these characteristic temperatures. Such a scheme
turns out to be fully equivalent to the general pattern of
magnetic anisotropy in (001) (Ga,Mn)As under compressive
strain.\cite{Wang:2005_PRL,Welp:2003_PRL,Tang:2003_PRL,Sawicki:2005_PRB,Wang:2005_PRB_a,Kato:2004_JJAP,Welp:2004_APL}
In such films uniaxial anisotropy between $[110]$ and
$[1\overline10]$ directions, dominating at elevated temperatures,
gives way at low $T$ to biaxial anisotropy with in-plane $\langle
100\rangle$ easy axes. This spin reorientation transition takes
place at a temperature, at which uniaxial and biaxial anisotropy
constants equilibrate,\cite{Wang:2005_PRL} and is corroborated
numerically in our samples from analysis of the magnetization
processes presented in Sec.~\ref{sec: phenomodel}.

In an analogy to (001) (Ga,Mn)As, let us assume for a moment that
$\bm M$ of a (113) sample remains (without a magnetic field) in
(001) plane. Then, a similar description in terms of two in-plane
anisotropies (one biaxial and one uniaxial) is possible.
Furthermore, assuming that the uniaxial anisotropy constant is
proportional to $M(T)^2$, the biaxial anisotropy constant is
proportional to $M(T)^4$ (Ref.~\onlinecite{Wang:2005_PRL}), and
that both equilibrate at $T^*$ we are able to model qualitatively
temperature induced rotation of magnetization in the sample and
calculate all three components of magnetization that would be
measured by SQUID. The thick solid lines in Fig.~\ref{fig:TRM}
show the results for sample S1 and we find them reproducing the
experimental findings reasonably well. Therefore we identify $T^*$
as the temperature at which the spin reorientation transition from
a biaxial anisotropy along $\langle 100 \rangle$ to uniaxial one
along $[1\overline10]$ takes place in this system. On the other
hand, the discrepancies seen in Fig.~\ref{fig:TRM} indicate that a
more elaborated model is needed. In particular, we can infer from
the low temperature TRM data that the orientation of $\bm M$ at
5~K moves actually away from  $\langle 100 \rangle$ on annealing.
The angle between $\bm M$ and  $\langle 100 \rangle$ is increasing
from 9, through 19 to 26~deg for samples S1, S2 and S3
respectively. At the same time the angle between $\bm M$ and (113)
plane is dropping from 13 to 7~deg. This indicates that the plane
in which both easy orientations of  $\bm M$ reside at low $T$ is
tilting away from (001) towards (113) plane. This observation is
fully confirmed from the comprehensive analysis of the magnetic
anisotropy presented in the next section.


We remark here that the origin of the symmetry breaking between
$[110]$ and $[1\overline10]$ in (001) (Ga,Mn)As is still unknown
and it is very stimulating to see favoring the in-plane
$[1\overline10]$ also in layers of different surface
reconstruction than (001) GaAs.

\subsection{Experimental determination of anisotropy constants}
\label{sec: phenomodel}

In order to build up a more complete anisotropy description we
analyze full magnetization curves $M(H)$. It was shown by Limmer
et al. (Ref.~\onlinecite{Limmer:2006_MJ,Limmer:2006_PRB}) that an
accurate description the magnetic anisotropy in (113)A (Ga,Mn)As
requires at least four components: a cubic magnetic anisotropy
with respect to the $\langle 001 \rangle$ axes, uniaxial in-plane
anisotropy along the $[\overline110]$ direction, and two uniaxial
out-of-plane anisotropies along the $[113]$ and $[001]$
directions. The first two anisotropy components are commonly
observed in (001)-oriented (Ga,Mn)As samples, and as shown in the
previous section they are sufficient to provide even
semi-quantitative description in (113) case. The other two arise
from the epitaxial strain and demagnetizing effect, both of which
depend on the orientation of the substrate. In our approach we
combine the two out-of-plane magnetic anisotropy contributions
into a single one, with its hard axis oriented between the $[001]$
and $[113]$ directions. Accordingly, we write the free energy in
the form,
\begin{eqnarray}
  \label{eq: F}
  F & = & -\mu_0\bm{H} \cdot \bm{M} + K_C (w_x^2 w_y^2 + w_y^2 w_z^2 + w_z^2 w_x^2 ) \nonumber\\
    & & {} + K_{u\overline110} \sin^2 \Theta \sin^2 \Phi \nonumber\\
    & & {} + K_{u1} (\cos\Theta_A \cos\Theta
          - \sin\Theta_A \sin\Theta \cos\Phi)^2.
\end{eqnarray}
Here, $K_C$, $K_{u\overline110}$ and $K_{u1}$ are the lowest order
cubic, in-plane uniaxial and out-of-plane uniaxial anisotropy
energies, respectively; $\Theta_A$ describes the angle between
the $K_{u1}$ hard axis and $[113]$ direction; $w_x$, $w_y$, and $w_z$ denote
direction cosines of the magnetization vector with respect to the main
crystallographic directions $\langle 100 \rangle$; $\Theta$ is the
angle between $\bm M$ and the $[113]$ direction, and $\Phi$ is the
angle between the projection of the $\bm M$ onto the sample plane and
the $[33\overline2]$ direction.

\begin{figure}[tp]
  \begin{center}
        \includegraphics[width=0.98\columnwidth]{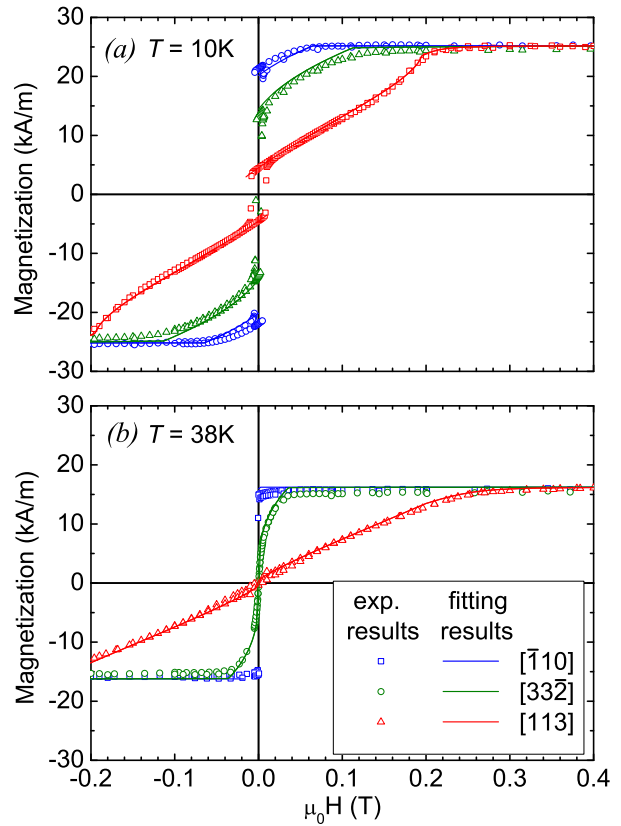}
  \end{center}
\caption{(Color online) Examples of magnetization curves for the
as-grown sample measured along three, mutually orthogonal, major
sample directions $[\overline110]$, $[33\overline2]$, and $[113]$:
(a) below spin reorientation transition at 10~K and (b) above, at
38~K. Symbols indicate measurement points; lines represent the
best fit of the model described by Eq.~(\ref{eq: F}).}
    \label{fig:M-H}
\end{figure}

By numerical minimizing of the free energy with respect to
$\Theta$ and $\Phi$ we are able to trace the rotation of $\bm M$,
starting from the given orientation, while sweeping or rotating
external magnetic field. Adjusting the obtained ``trace'' to the
experimental data we get the values of the four parameters of the
model. We perform this procedure numerically for every sample for
all three orientations and for all temperatures the $M_i(H)$
curves have been recorded. Figure~\ref{fig:M-H} shows an example
of the measured and fitted $M_i(H)$ for the sample S1 at two
different temperatures.

\begin{figure*}[tbp]
  \begin{center}
        \includegraphics[width=0.98\textwidth]{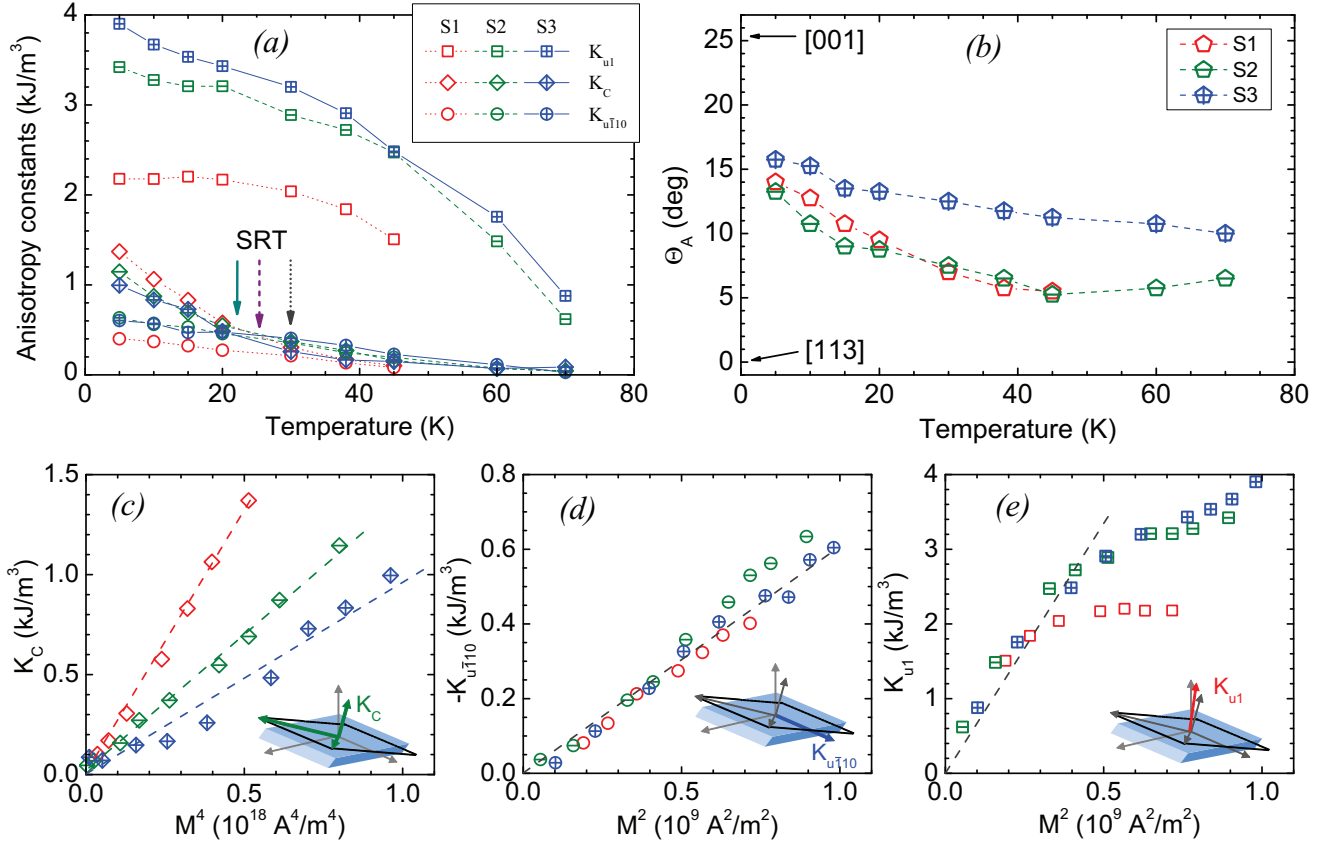}
  \end{center}
\caption{(Color online) (a) and (b) points: temperature dependence
of $K_C$, $K_{u\overline110}$ and $K_{u1}$ (a) and angle
$\Theta_A$ (b) obtained from numerical fitting of Eq.~(\ref{eq:
F}) to experimental magnetization curves for all three samples
considered in this study. Solid, dashed and dotted arrows in (a)
indicate the spin reorientation temperature in samples S1, S2 and
S3, respectively. (c), (d) and (e): $K_C$, $K_{u\overline110}$ and
$K_{u1}$ dependence on $M^4$, $M^2$ and $M^2$, respectively. In
all panels, the various lines are guides for eye only.}
  \label{fig:magnetic anisotropyconstants}
\end{figure*}

\begin{figure}[tbp]
  \begin{center}
       \includegraphics[width=0.95\columnwidth]{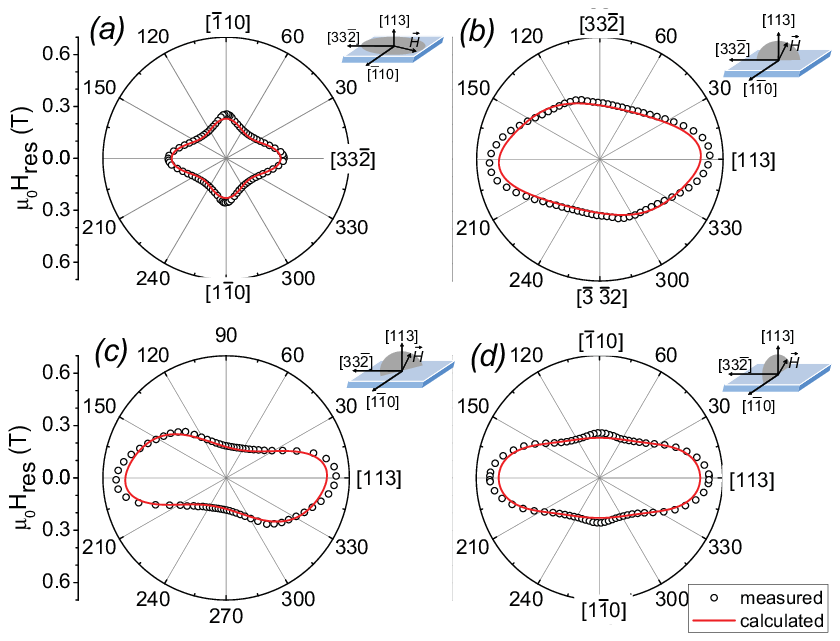}
  \end{center}
\caption{(Color online) Angular dependence of the ferromagnetic
resonance fields for external magnetic field rotating in four
different crystallographic planes: $(113)$, $(\overline{1}10)$,
$(33\overline{2})$ and $\frac{1}{2\sqrt{11}}(3-\sqrt{11},
3+\sqrt{11}, -2)$. Points show measured resonance field values.
Lines show resonant field values calculated using anisotropy
energy described by Eq.~(\ref{eq: F}) and magnetic anisotropy
constants obtained from hysteresis loops.}
  \label{fig: 5}
\end{figure}

Temperature dependence of the three magnetic anisotropy constants
and the angle $\Theta_A$ are presented in Fig.~\ref{fig:magnetic
anisotropyconstants}a and b, respectively. All $K_i$'s
monotonically decrease with temperature, and, like in (001)
(Ga,Mn)As, the cubic anisotropy constant $K_C$ (with $\langle 100
\rangle$ easy axes, see Fig.~\ref{fig:magnetic
anisotropyconstants}c) and in-plane uniaxial constant
$K_{u\overline110}$ (with $[1\overline10]$ easy axis, see
Fig.~\ref{fig:magnetic anisotropyconstants}d) are proportional to
$M^4$ and $M^2$, respectively, so confirming the validity of the
single domain approach used to analyze the observed magnetization
rotations. The $K_C(T)$ and $K_{u\overline110}(T)$ data point to
the presence of the spin reorientation transition in (001) plane.
This already inferred from TRM data magnetic easy axis changeover
must take place as $K_C(T)$ and $K_{u\overline110}(T)$ swap their
intensities in our samples. The relevant temperatures are marked
in Fig.~\ref{fig:magnetic anisotropyconstants}a by arrows.
Importantly, we find these temperatures to agree within 2-3~K with
those indicated in Fig.~\ref{fig:TRM}, what strongly underlines
the correctness of the approach we employ here to describe the
magnetic anisotropy in our samples. We note that the SRT shifts
to lower temperatures on going from sample S1 to S3, since on
annealing the in-plane uniaxial anisotropy gets strongly enhanced
relative to the cubic one (compare Figs.~\ref{fig:magnetic
anisotropyconstants}c and d).

In contrast, $K_{u1}$ shows a more complex dependence on $M^2$,
see Fig.~\ref{fig:magnetic anisotropyconstants}e. A
proportionality of the out-of-plane anisotropy constant to $M^2$
is seen only at low $M$, that is at high $T$. On lowering
temperature $K_{u1}$ departures from this trend, and the effect is
the strongest for the sample~S1. We connect this behavior with
the proximity of this system to another spin reorientation
transition, the transition from the hard to easy out-of-plane axis
of the $K_{u1}$ uniaxial magnetic anisotropy. Such a SRT takes
place in compressively strained (001) (Ga,Mn)As on lowering $T$,
and was already observed in samples with moderate or high $x$ but
rather low hole density\cite{Sawicki:2004_PRB}. The effect depends
on the ratio of valence band spin splitting to the Fermi energy.
Therefore the S1 sample, the one with the lowest $p$ is expected
to show the strongest deviations from the expected functional
form. Then on annealing, along with the increase of $p$, we expect
the so called in-plane magnetic anisotropy (for the compressively
strained layers) to become more robust [less dependent on the
magnitude of the valence band splitting, that is on $M(T)$], as
experimentally observed.

Finally, we comment on $\Theta_A$, the parameter that can serve as
a measure of the angle between an `easy plane' with respect to
$K_{u1}$ (perpendicular) hard axis and the sample face. As
indicated in Fig.~\ref{fig:magnetic anisotropyconstants}b that
angle remains nearly constant at elevated temperatures and shows a
weak, but noticeable turn towards [001] below temperatures which
can be associated with $K_{u\overline110} \Leftrightarrow K_C$
SRT. This behavior again indicates the departure of easy direction
of $\bm{M}$ from $[1\overline10]$ direction in the (113) plane.
However, the maximum determined value of $\Theta_A \cong 15^o$
indicates, that the rotation of $\bm{M}$ actually neither takes
place in the (001) plane, nor is it directed exactly towards
$\langle 100 \rangle$ directions. $\bm{M}$ rather follows a
complex route in between (001) and (113) planes, a conclusion that
is a numerical confirmation of the results of the simple analysis
of the TRM data presented in the previous section.

\subsection{Ferromagnetic resonance}
\label{sec: fmr}

A tool widely used to study magnetic anisotropy is  ferromagnetic
resonance spectroscopy. Magnetic anisotropy in thin (Ga,Mn)As
films on (113)A GaAs was recently studied by
Bihler\cite{Bihler:2006_APL} and
Limmer\cite{Limmer:2006_MJ,Limmer:2006_PRB}. In a ferromagnetic
resonance experiment the magnetization vector $\bm M$ of the
sample precesses around its equilibrium direction in given
external magnetic field $\bm H$ with Larmor frequency $\omega_L$.
The resonant condition at a fixed frequency of microwaves $\omega$
is given by,
\begin{equation}
  \label{eq: 2}
  \left( \frac{\omega}{\gamma} \right)^2 =
    \frac{1}{\sin^2\Theta} \left[ \frac{\partial^2 F}{\partial
      \Theta^2} \frac{\partial^2 F}{\partial \Phi^2}
      - \left( \frac{\partial^2 F}{\partial \Theta \, \partial
      \Phi} \right)^2 \right].
\end{equation}
Here, $\gamma = g \mu_B \hbar^{-1}$ is the gyromagnetic ratio,
$g$ is  the $g$-factor, $\mu_B$ the Bohr magneton, and $\hbar$ is
the Planck constant. Resonance field is obtained by evaluating
Eq.~(\ref{eq: 2}) at the equilibrium position of $\bm M$
($\partial F/\partial \Theta = 0$ and $\partial F/\partial \Phi =
0$).

In Fig.~5 the dependence of the measured resonant  fields on the
orientation of the applied magnetic field is shown for the sample
S2, along with the results of a calculation made according to
Eq.~(\ref{eq: 2}) with the magnetic anisotropy parameters obtained
from SQUID magnetization curves. The agreement between the
calculation and the measured data is very good, indicating that
Eq.~(\ref{eq: F}) captures main features of  magnetic anisotropy
and that the numerical procedure employed to extract the
anisotropy constants is correct.

\subsection{Magnetization reversal}

\begin{figure}[tbp]
  \begin{center}
       \includegraphics[width=0.9\columnwidth]{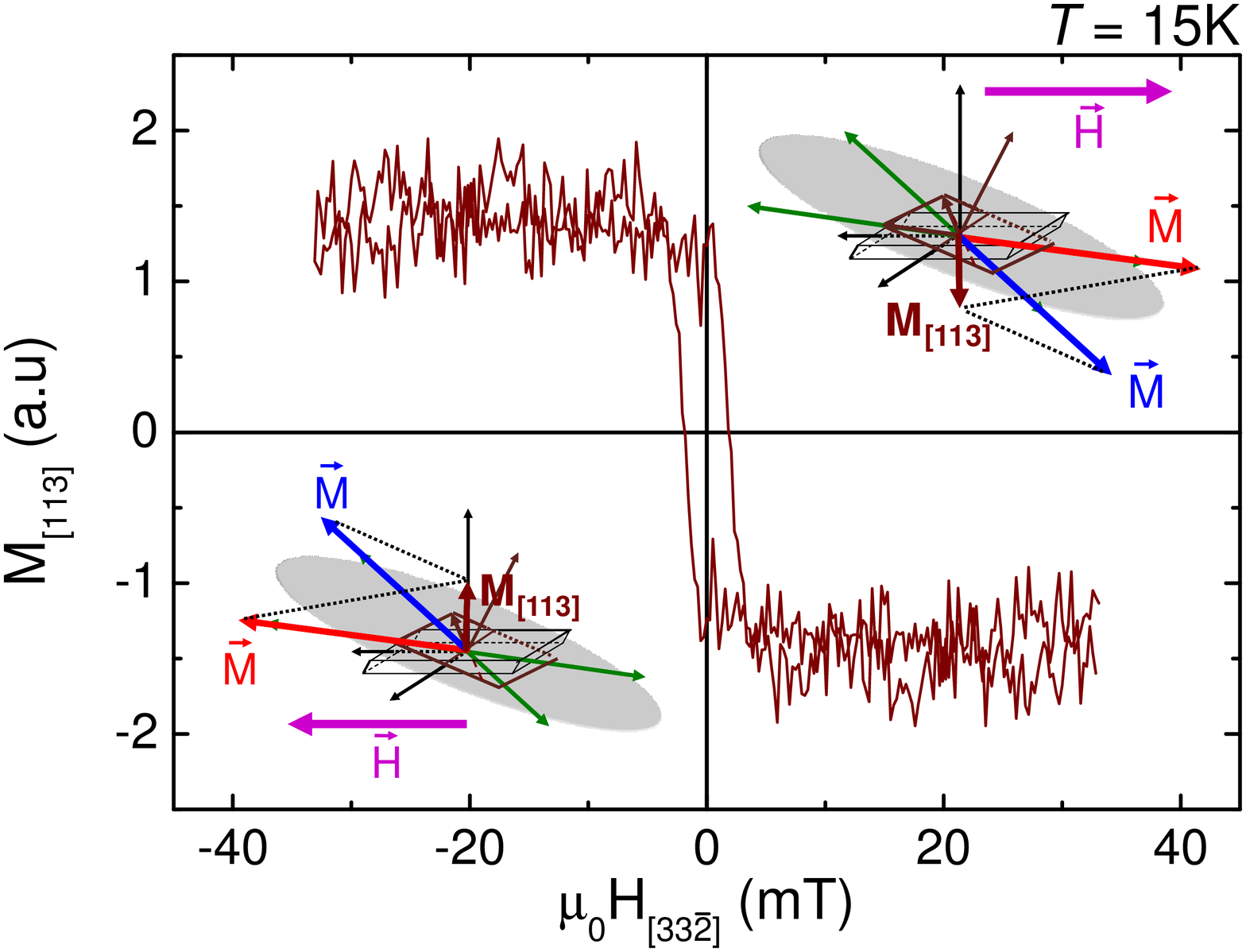}
  \end{center}
\caption{(Color online) The out-of-plane magnetization component
dependence on the in-plane magnetic field (swept along
$[33\overline2]$). The measurement is performed on the as-grown
sample S1 at $T = 15$~K. The cartoons inserted in the figure
illustrate the process.}
  \label{fig: 06}
\end{figure}

We end the experimental part evidencing an interesting mechanism
of the reversal of the out-of-plane magnetization component by an
in-plane magnetic field.

Below the spin reorientation transition (about 20-30~K) the
magnetization easy axes are moving close to the $\langle 100
\rangle$ directions, that is they are tilted up from the sample
face, the (113) plane, and so $\bm M$ is acquiring a sizable
non-zero component $M_{[113]}$. These two axes define a plane
lying between (113) and (001) planes, which shares the common
$[\overline110]$ direction with those two planes. Therefore, any
sweep of an external field, except that along the $[\overline110]$
direction, will result in a magnetization rotation across the
$[\overline110]$ line from one half of that plane (say that one
`above' sample face) to the other one (say `below') resulting in
$M_{[113]}$ reversal.

Such a process is illustrated in Fig.~\ref{fig: 06} in the most
interesting case, when the field is swept in the sample plane,
along $[33\overline2]$. We record the $M_{[113]}$ magnetization
component using the polar MOKE technique and a clear change of
sign of the signal evidences the reversal of $\bm M$ by the
application of an in-plane magnetic field. The cartoons inserted
in this figure visualize the mechanism of this reversal.

\section{Theory of magnetic anisotropy}
\label{sec: theory}
\subsection{$\bm{k} \cdot \bm{p}$ Hamiltonian}
The current theory describing the properties of the  (Ga,Mn)As
ferromagnetic semiconductor is the $p$-$d$ Zener
model.\cite{Dietl:2000_S} In this model, the thermodynamic
properties are determined by the valence band carriers
contribution to the free energy of the system, which is calculated
taking the spin-orbit interaction into account within the $\bm{k}
\cdot \bm{p}$
theory\cite{Dietl:2000_S,Dietl:2001_PRB,Jungwirth:2006_RMP,Zemen:2009_PRB}
or tight binding model\cite{Werpachowska:2009_arXiv} with the
$p$-$d$ exchange interaction between the carriers and the
localized Mn spins considered within the virtual-crystal and
molecular-field approximations.  Within this approach, magnetic
anisotropy depends on the strain tensor components.

The 6-band Luttinger $\bm{k} \cdot \bm{p}$ Hamiltonian of
a valence band electron in a zinc-blende semiconductor is a block
matrix (cf.\ Ref.~\onlinecite{Sliwa:2006_PRB}):
\begin{equation}
  \mathcal{H}_{\mathrm{6x6}} = \begin{pmatrix}
    H^{vv}& H^{vs} \\
    H^{sv}& H^{ss}
  \end{pmatrix},
\end{equation}
where
\begin{eqnarray}
  \mathcal{H}^{vv} & = & - \frac{\hbar^2}{m} \left\{
    \frac12 \gamma_1 k^2
    - \gamma_2 \left[\left(J_x^2 - \frac13 J^2\right) k_x^2
      + c.p. \right] \right. \\
  && \left. {} - 2 \gamma_3 \left[
    \{J_x, J_y\} \{k_x, k_y\} + c.p. \right] \right\} \nonumber \\
  \mathcal{H}^{vs} & = & - \frac{\hbar^2}{m} \left[
    -3\gamma_2 (U_{xx} k_x^2 + c.p.) \right. \\
  && \left. {} - 6 \gamma_3 (U_{xy} \{k_x, k_y\} + c.p.) \right]
    \nonumber \\
  \mathcal{H}^{ss} & = &
    - \left( \Delta_0 + \frac{\hbar^2}{2 m} \gamma_1 k^2 \right)
\end{eqnarray}
(we use the notation of Ref.~\onlinecite{Sliwa:2006_PRB}). Our basis is related to that of Ref.~\onlinecite{Dietl:2001_PRB} as follows: $u_1 = -\left| \frac32, \frac 32 \right>$, $u_2 = -i\left| \frac32, \frac 12 \right>$, $u_3 = \left| \frac32, -\frac 12 \right>$, $u_4 = i\left| \frac32, -\frac 32 \right>$, $u_5 = -\left| \frac12, \frac 12 \right>$, $u_6 = i\left| \frac12, -\frac 12 \right>$, \emph{i.e.} we use the standard basis of angular momentum eigenvectors (notice the change of sign in $\left| \frac12, \frac 12 \right>$ and $\left| \frac12, -\frac 12 \right>$ with respect to Ref.~\onlinecite{Sliwa:2006_PRB} that accounts for the difference in the sign of $H^{vs}$, $H^{sv}$). In this basis the $p-d$ exchange Hamiltonian is:
\begin{equation}
  \mathcal{H}_{pd} = B_G \begin{pmatrix}
   2 ( \bm{J} \cdot \bm{w} )&
  -6 ( \bm{U} \cdot \bm{w} )\\
  -6 ( \bm{T} \cdot \bm{w} )&
  - ( \bm{\sigma} \cdot \bm{w} )
  \end{pmatrix},
\end{equation}
where $\bm{w} = \bm{M}/M$ and $B_G$
is given by equation~2 of Ref.~\onlinecite{Dietl:2001_PRB},
\begin{equation}
  B_G = A_F \beta M /(6 g \mu_B),
  \label{eq:B_G}
\end{equation}
while the strain Hamiltonian is:
\begin{eqnarray}
  \mathcal{H}^{vv}_{\epsilon} & = & -b \Bigl[ (J_x^2 - \frac13 J^2) \epsilon_{xx} +
  c.p. \Bigr] \\ & & {} - \frac{d}{\sqrt{3}} \Bigl[ 2\{ J_x, J_y \}
  \epsilon_{xy} + c.p. \Bigr] \\
  \mathcal{H}^{vs}_{\epsilon} & = & -3 b (U_{xx} \epsilon_{xx} + c.p.) \\
  & & {} - \sqrt{3} d (2 U_{xy} \epsilon_{xy} + c.p.) \\
  \mathcal{H}^{ss}_{\epsilon} & = & 0.
\end{eqnarray}

Since the strain tensor for the $(113)$ substrate orientation features non-zero non-diagonal components, it is necessary to include in the $\bm{k} \cdot \bm{p}$ Hamiltonian the so called $k$-linear terms, {\em i.e.}, terms linear in $\bm{k}$ and $\bm{\epsilon}$ coming \emph{via} second-order perturbation (Ref.~\onlinecite{Bir:1974_B}, \S 15) from the terms in the $8\times 8$ Kane Hamiltonian\cite{Pikus:1984_B} that mix the conduction and the valence bands. The corresponding Hamiltonian is
\begin{equation}
  \mathcal{H}_{k\epsilon} = C_4 \begin{pmatrix}
    \bm{J} \cdot \bm{\varphi}&
      \frac32 \left( 1 - \frac\eta 2 \right)
      \bm{U} \cdot \bm{\varphi}\\
    \frac32 \left( 1 - \frac\eta 2 \right)
      \bm{T} \cdot \bm{\varphi}&
    \left( 1 - \eta \right) \bm{\sigma} \cdot \bm{\varphi}
  \end{pmatrix},
\end{equation}
where $\eta = \Delta_0/(E_g + \Delta_0)$ and the components of the vector $\bm{\varphi}$ are $\varphi_z = \epsilon_{zx} k_x - \epsilon_{zy} k_y$ (c.p.). The numerical value given in
Ref.~\onlinecite{Pikus:1984_B} is $C_3/\hbar = 8 \times 10^5 \, \mathrm{m} \, \mathrm{s}^{-1}$, hence for $C_4 = -C_3/(2 \eta)$ we obtain $C_4 / \hbar = -2.2 \times 10^6 \, \mathrm{m/s}$.

\subsection{Strain tensor}

Determining the components of the strain tensor for an unrelaxed epitaxial layer grown on a lattice mismatched substrate can be considered a classical topic. The two possible approaches to this problem are (i) to solve a system of linear equations for the strain and stress components assuming that some components of those tensors vanish (this is our approach) or (ii) to determine the strain of the layer by minimizing the elastic energy (this is the approach formulated in Ref.~\onlinecite{Yang:1994_APL}). Our approach involves a transformation of the coordinate system that is feasible in general only using a computer algebra system. Using one we arrive to the form of the symmetric strain tensor that is in a perfect agreement with that of Ref.~\onlinecite{Yang:1994_APL}.

The result for a $(11n)$ oriented substrate is,
\begin{equation}
  \bm{\epsilon}_{\text{epi}} = \frac{f}{A+B+C}
    \begin{pmatrix}
          B+C& -A& -D/2\\
          -A& B+C& -D/2\\
          -D/2& -D/2& B-2C
    \end{pmatrix},
\end{equation}
where
\begin{eqnarray}
  A & = & 3\left[n^2(c_{11}-c_{12})-(n^2-2)c_{44}\right] (c_{11}+2c_{12}), \\
  B & = & 2 \left[(n^4+4n^2+1) (c_{11}-c_{12})+6 c_{44}\right] c_{44}, \\
  C & = & (n^2-1)(n^2+2) (c_{11}+2c_{12})c_{44}, \\
  D & = & 3n \left[(n^2+1)(c_{11}-c_{12})+2 c_{44}\right] (c_{11}+2c_{12}),\quad
\end{eqnarray}
and
\begin{equation}
  f = -\Delta a/a = (a_0-a)/a
\end{equation}
is the relative lattice constant misfit.
Here, the components of the epitaxial strain tensor are given with respect to the coordinates $(x, y, z)$ associated with the crystallographic axes, $x = [100]$, $y = [010]$, and $z = [001]$ (\emph{i.e.} the quantities given above are the strain components that enter the $\bm{k} \cdot \bm{p}$ Hamiltonian).

Using the values $c_{11} = 119. \, \mathrm{GPa}$, $c_{12} = 53.8 \, \mathrm{GPa}$, and $c_{44} = 59.5 \, \mathrm{GPa}$ (Ref.~\onlinecite{Madelung:1991_B}, p.~105) we obtain
\begin{equation}
  \bm{\epsilon}_{\text{epi}} = -\frac{\Delta a}{a}
    \begin{pmatrix}
      \phantom{-}0.9488 & -0.0512 & -0.3478 \\
      -0.0512 & \phantom{-}0.9488 & -0.3478 \\
      -0.3478 & -0.3478 & -0.6260
    \end{pmatrix}.
\end{equation}

For the purpose of determining the strain components from X-ray diffraction data, the components of the strain tensor in the coordinate system associated with the epitaxial film are needed. We take as the coordinate system: $x' = [3, 3, -2]$, $y' = [-1, 1, 0]$, $z' = [1, 1, 3]$. The relative difference of the lattice constants along the $[113]$ direction between that layer and the substrate is
\begin{widetext}
\begin{eqnarray}
\label{eq:Delta_d}
\Delta d/d &= & \frac{[27(c_{11}-c_{12})+67c_{44}](c_{11}+2c_{12})}{9 (c_{11} - c_{12})(c_{11} + 2 c_{12}) + (101 c_{11}-34c_{12}+4c_{44}) c_{44}} \Delta a/a =  1.7284 \, \Delta a/a.
\end{eqnarray}
For the sake of completeness we notice that there is also a shear strain component
\begin{eqnarray}
\epsilon_{x'z'} &=& \frac{12\sqrt{2}[(c_{11}-c_{12})-2c_{44}](c_{11}+2c_{12})}{9 (c_{11} - c_{12})(c_{11} + 2 c_{12})+(101c_{11}-34c_{12}+4c_{44}) c_{44}}\Delta a/a = -0.2746 \, \Delta a/a.
\end{eqnarray}
\end{widetext}

Following Ref.~\onlinecite{Sawicki:2005_PRB},  to account for the
mechanism which generates the in-plane uniaxial anisotropy in
$(001)$ samples, we incorporate in the $p-d$ Zener model an
additional Hamiltonian term corresponding to shear strain
$\bm{\epsilon}'$,
\begin{equation}
  \bm{\epsilon} = \bm{\epsilon}_{\text{epi}} + \bm{\epsilon}'.
\end{equation}
In case of a $(001)$-oriented substrate the additional strain $\bm{\epsilon}'$ has a non-zero $xy$ component, $\epsilon'_{xy}$. The corresponding anisotropy is of the form $K_{xy} w_x w_y$ (as in Ref.~\onlinecite{Bowden:2008_JPCM}), hence it is a difference of uniaxial anisotropies on the $[110]$ and $[\overline110]$ directions, and the anisotropy field is $H_u = 2 K_{xy}/(\mu_0 M)$ (this is the field required to align magnetization along the hard axis, e.g. $[110]$; only $H_u/2$ is required to align magnetization along the $z$ direction). In the case of a $(113)$-oriented substrate the additional strain may have more non-zero components. We assume that the mirror symmetry with respect to the $(\overline110)$ plane is preserved, hence $\epsilon'_{xz} = \epsilon'_{yz}$.

\subsection{Numerical procedure}
The numerical procedure serving to determine the magnetic anisotropy from the Hamiltonian matrix is described in Ref.~\onlinecite{Dietl:2001_PRB}. Let us note that including the $k$-linear terms in the Hamiltonian leads to a tenfold increase of the processing time, although in specific cases it is possible to generate a symbolic expression for the characteristic polynomial of the $6\times 6$ Hamiltonian matrix. Moreover, since numerical interpolation of the dependence of the hole concentration on the Fermi energy may lead to uncontrollable inaccuracies, an alternative procedure that avoids those inaccuracies is to directly integrate the energy of the carriers in the momentum space. However, the integration has to be done separately for each hole concentration (this is an advantage if a single hole concentration is specified). Moreover, one still needs to solve the inverse eigenvalue problem to find the discontinuities of the integrand.

In a numerical calculation, it is possible to determine the full magnetic anisotropy by computing the free energy of the carriers for a number of directions of magnetization. In our case we choose a grid of directions that is rectangular in the spherical coordinates ($w_x = \sin\theta\cos\phi$, $w_y = \sin\theta\sin\phi$, $w_z = \cos\theta$), \emph{i.e.} $\theta = \theta_i$ and $\phi = \phi_j$, where $\cos\theta_i$, $i = 1, 2, \ldots, N_\theta$ are the nodes of a Gaussian quadrature and $\phi_j = 2 \pi j / N_\phi$, $j = 0, 1, \ldots, N_\phi-1$ are equally spaced. Then, following the method used in the software package SHTOOLS\cite{Wieczorek:2009_WWW} (routine \texttt{SHExpandGLQ}) we expand the magnetic anisotropy (free energy) into a sum of low-order spherical harmonics. Since the free energy is even, choosing even $N_\phi$ allows to restrict the grid to a half of the sphere. We use the standard quantum mechanics (orthonormalized) spherical harmonics $Y_{lm}(\theta, \phi)$, and denote the coefficients of this expansion $r_{lm}, m = 0, 1, \ldots, l$, and $s_{lm}, m = 1, 2, \ldots l$,
\begin{equation}
\label{eq:F_theory}
  F_c = \sum_{l} \left[ r_{l0} Y_{l0} + \sum_{m = 1}^{l} (r_{lm} \Re Y_{lm} + s_{lm} \Im Y_{lm}) \right],
\end{equation}
where the outer sum is over $l = 0, 2, \ldots, N_{\theta}-1$. The scalar product in this representation is diagonal, with a weight of $1$ for $r_{l0}$ and $1/2$ for $r_{lm}$, $s_{lm}$, $m > 0$.

\subsection{Magnetic anisotropy}
There are a few sources of magnetic anisotropy in epitaxial $\mathrm{Ga}_{1-x}\mathrm{Mn}_{x}\mathrm{As}$: the cubic anisotropy of the valence band, epitaxial strain, the additional off-diagonal strain $\epsilon'$, and the shape anisotropy caused by the demagnetization effect. Since $\epsilon'$ is unknown, it is inevitable to parametrize the anisotropy in a manner that separates the components affected by $\epsilon'$ from what is predictable. If we measure $(\theta, \phi)$ in the spherical harmonic representation of the carriers' free energy with respect to the crystallographic axes, the above spherical harmonic representation allows one to separate the components brought about by  a non-zero value of $\epsilon'$ from the remaining sources of magnetic anisotropy. Indeed, as far as $l = 2$ is concerned, $\epsilon'$ affects primarily only $K_{xy}$, $K_{yz}$, and $K_{xz}$, where $K_{xy} = \sqrt{\frac{15}{8\pi}} s_{22}$, $K_{yz} = -\sqrt{\frac{15}{8\pi}} s_{21}$, and $K_{xz} = -\sqrt{\frac{15}{8\pi}} r_{21}$. The remaining components are $r_{20}$ and $r_{22}$. As $r_{22} = 0$ due to the mirror symmetry, they can be collected into one term $K_{u001} (w_z^2 - \frac13)$, with $K_{u001} = \frac34 \sqrt{\frac{5}{\pi}} r_{20}$. Thus, we describe the $l = 2$ anisotropy (without the demagnetization contribution) by $K_{xy}$, $K_{xz} = K_{yz}$, and $K_{u001}$. Finally, the cubic anisotropy corresponds to $(r_{40}, r_{44}) = -\frac{2 \sqrt{\pi}}{15} K_C (1, \sqrt{10/7})$.

We have to relate now the components of the spherical harmonic
representation, $r_{lm}$ and $s_{lm}$, to the experimentally
determined magnetic anisotropy constants $K_C, K_{u1},
K_{u\overline110} $, and $\Theta_A$, as specified in Eq.~\ref{eq: F}
and presented in Fig.~\ref{fig:magnetic anisotropyconstants}.
Since the demagnetization effect adds to
$(r_{20}, r_{21}, s_{21}, r_{22}, s_{22})$ the contribution
$\frac{4\sqrt{10 \pi}}{165} K_d (4 \sqrt2, -3 \sqrt3, -3 \sqrt3, 0, \sqrt3)$,
with $K_d = \mu_0 M^2/2$, the constants are related
to those of Eq.~(\ref{eq: F}) as follows,
\begin{eqnarray}
  K_{u001} & = & \frac{1 + 3 \cos 2\Theta_A'}4 K_{u1} - \frac12 K_{u\overline110} - \frac{8}{11} K_d, \label{eq:Ku001} \quad \\
  K_{xy} & = & \frac{1 - \cos 2\Theta_A'}2 K_{u1} - K_{u\overline110} - \frac2{11} K_d, \label{eq:Kxy} \\
  K_{xz} & = & -\frac{\sqrt2 \sin 2\Theta_A'}{2} K_{u1} - \frac{6}{11} K_d, \label{eq:Kxz}
\end{eqnarray}
where $\Theta_A' = \Theta_A - \arccos(3/\sqrt{11})$.


\begin{figure}[b]
  \begin{center}
       \includegraphics[width=0.95\columnwidth]{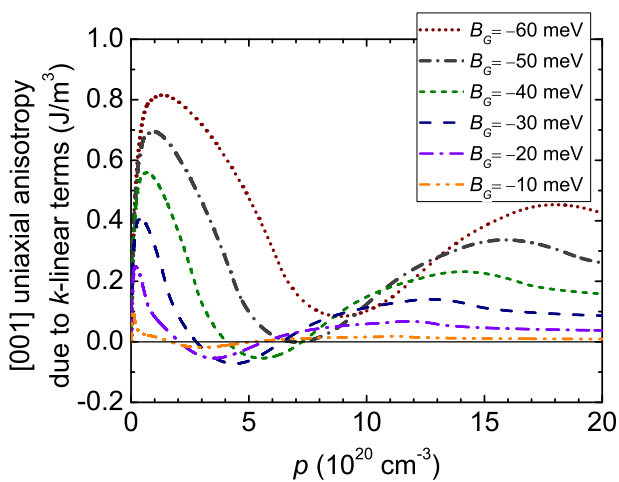}
  \end{center}
  \caption{(Color online) The contribution of the $k$-linear terms to the $[001]$ uniaxial anisotropy for $C_4 / \hbar = -2.18 \times 10^6 \, \mathrm{m/s}$ and  $\epsilon_{xy} = 0.05\%$.}
  \label{fig: 7}
\end{figure}

\begin{figure}[b]
  \begin{center}
       \includegraphics[width=0.95\columnwidth]{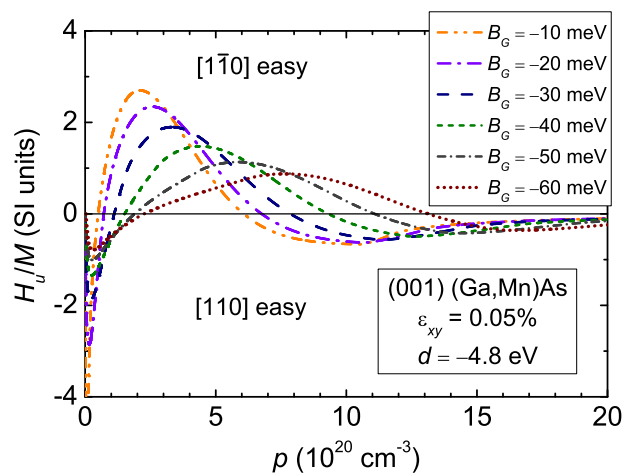}
  \end{center}
\caption{(Color online) Hole concentration dependence of the
in-plane uniaxial anisotropy field due to shear strain
$\epsilon_{xy} = 0.05\%$ for various values of the valence-band
spin-splitting.}
  \label{fig: 8}
\end{figure}

We carry out numerical calculations with band structure
parameters and deformation potentials specified
previously.\cite{Dietl:2001_PRB,Sawicki:2005_PRB} We include
hole-hole exchange interactions {\em via} the Landau parameter of
the susceptibility enhancement,  $A_F = 1.2$
(Ref.~\onlinecite{Dietl:2001_PRB}). This parameter, assumed here
to be independent of the hole density and strain, enters into the
relation between $M$ and $B_G$, but also divides the anisotropy
constants. More specifically, we make the calculation with $B_G$
enhanced by the factor $A_F$, and divide the resulting anisotropy
constants by $A_F^{n-1}$, where $n$ is the power of
magnetization~$M$ to which a given anisotropy constant is
proportional. The result is proportional to $A_F$. We have $n = 2$
for the uniaxial anisotropies and $n = 4$ for the lowest order
cubic anisotropy (the proportionality holds for $B_G$ smaller than
a few meV). We note that the cubic anisotropy field shown in
Fig.~9 of Ref.~\onlinecite{Dietl:2001_PRB} was divided by $A_F$
rather than $A_F^3$.

To evaluate the effect of the $k$-linear terms,  we use a non-zero
value of $C_4$ and calculate the difference of the resulting
anisotropy with respect to the $C_4 = 0$ case. This difference has
only one noticeable component, $\Delta r_{20} = r_{20}(C_4 \ne
0)-r_{20}(C_4 = 0)$, which corresponds to a uniaxial anisotropy
with a $[001]$ axis (or with $[100]$, $[010]$ axes for
$\epsilon_{yz} \ne 0$, $\epsilon_{xz} \ne 0$ respectively). A plot
of $\Delta K_{u001} \propto \Delta r_{20}$ is shown in
Fig.~\ref{fig: 7} for $\epsilon_{xy} = 0.05\%$ (as implied by the
symmetry, $\Delta K_{u001}$ is second order in $\epsilon_{xy}$).
The values are rather small. In fact, assuming $\Delta a/a =
0.5\%$, we have $\epsilon_{xz} = \epsilon_{yz} \approx 0.17\%$,
and the magnitude of $\Delta K_{u100} = \Delta K_{u010}$ is below
$10 \, \mathrm{J}/\mathrm{m}^3$ if we consider the epitaxial
strain only. This estimate appears to remain valid in case of a
general strain of a similar magnitude, although other anisotropy
components are affected as well and the dependence on strain
components is non-linear. However, we stress that this estimate
depends on the value of the parameter $C_4$, which is somewhat
uncertain and may be different for the ordinary strain and
$\bm{\epsilon}'$. Considered this, it is justified to set $C_4 =
0$ in the remaining part of this paper.

Before we proceed to the calculations specific to the particular
samples, we make a remark that the data originally shown in Fig.~6
of Ref.~\onlinecite{Sawicki:2005_PRB} were not correct due to a
numerical error in the form of the strain Hamiltonian. We show
corrected results for~$H_u$ in Fig.~\ref{fig: 8}. The present
results are in agreement with Fig.~17 of
Ref.~\onlinecite{Zemen:2009_PRB} (remember that our model includes
the parameter the Landau parameter $A_F$, neglected in
Ref.~\onlinecite{Zemen:2009_PRB}).

As discussed previously,\cite{Sawicki:2004_PRB,Sawicki:2005_PRB}
owing to sign oscillations of anisotropy constants, the direction
of magnetization can be changed by temperature ($B_{\text{G}}$) or
the hole concentration, particularly in the vicinity of $p =
6\times 10^{20}$ and $1\times 10^{20}$~cm$^{-3}$, according to the
results displayed in Fig.~8  The corresponding in-plane spin
reorientation transition has indeed been observed by some of us
either as a function of temperature\cite{Sawicki:2005_PRB} or the
gate voltage in metal-insulator semiconductor
structures\cite{Chiba:2008_N,Sawicki:2010_NP} in these two hole
concentration regions in (Ga,Mn)As, respectively.

\section{Comparison between experiment and theory}

We detailed above a microscopic model of the magnetic anisotropy
in a DMS. In order to assess the applicability of this model to an
arbitrarily oriented DMS we compare its predictions with
experimental findings for (113) (Ga,Mn)As.

First, we specify the magnitude of the lattice mismatch to
establish the components of the strain tensor. According to Fig.~2
of Ref.~\onlinecite{Daeubler:2006_APL}, for a $(113)$ sample
containing 6.4\% of Mn we expect $\Delta d/d = 5.6 \times 10^{-3}$
which, employing Eq.~\ref{eq:Delta_d}, translates into $\Delta a/a
= 0.323\%$. We assume this value throughout this section.

Then, using the values of $x_{\text{eff}}$ already established in
Sec.~\ref{seq: samples}, we calculate
\begin{equation}
  M_{\text{Mn}} = x_{\text{eff}} N_0 S g \mu_B
  \label{eq: M_Mn}
\end{equation}
and obtain $B_G$ for each of our samples from Eq.~\ref{eq:B_G}. It
is worth repeating here, that the established upon total $x$ and
$M_{\text{exp}}$ values of $x_{\text{eff}}$ and $p$ reproduce,
within the same model, experimental values of $T_{\text{C}}$
remarkably well. This boosts our confidence in the accuracy of the
material parameters used here for the computations of the magnetic
anisotropy and gives a solid ground to the presented conclusions.

\begin{figure*}[htp]
  \begin{center}
      \includegraphics[width=0.95\textwidth]{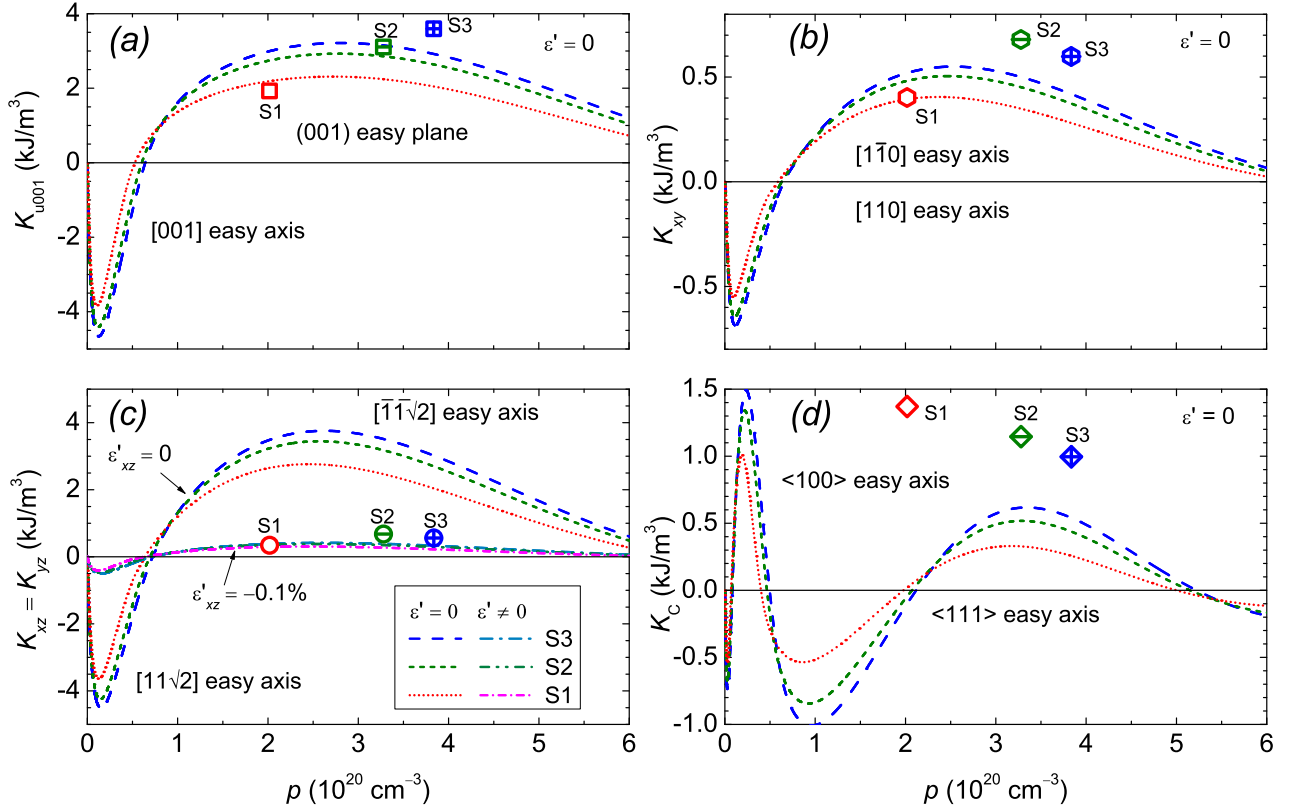}
  \end{center}
\caption{(Color online) Lines: theoretical dependence of the
anisotropy constants (a) $K_{u001}$, (b) $K_{xy}$, (c) $K_{xz}$,
and (d) $K_C$ on the hole concentration~$p$, calculated within
mean-field Zener model. The values of the exchange parameter $B_G$
and the lattice constant mismatch $\Delta a/a$ are specific to the
investigated samples and are specified in the text.  Symbols
depict values determined from the experiment. $\bm{\epsilon}' = 0$
is assumed here except for (c), where also the case of
$\epsilon'_{xz} = -0.1\%$ is included.}
  \label{fig:All_K}
\end{figure*}

The calculations are performed as a function of $p$, employing for
each sample the corresponding value of $B_G$: $-16.4$, $-18.7$, and
$-19.7$~meV for samples S1, S2, and S3, respectively. The results
are presented in Fig.~\ref{fig:All_K} as curves, while full
symbols represent experimentally established values of the
anisotropy constants (a square for the sample S1, a triangle for
S2, and a circle for S3). The experimental $K_{u001}$, $K_{xy}$,
$K_{xz}$, and $K_C$ are obtained from $K_{u\overline110}$,
$K_{u1}$, $K_C$ and $\Theta_A$ using Eqs.~\ref{eq:Ku001},
\ref{eq:Kxy} and \ref{eq:Kxz}. We take the $T = 5 \, \mathrm{K}$
experimental anisotropy data, as this is what is consistent with
the $T = 0$ limit, implicitly assumed in Eq.~\ref{eq: M_Mn}.

We start by discussing the strongest component of the magnetic
anisotropy, the $K_{u001}$ term. The calculated curves are
presented in Fig.~\ref{fig:All_K}a. The calculations have been
performed without introducing the fictitious shear strain
$\bm{\epsilon}'$ (\emph{i.e.} $\epsilon' = 0$). However, since we know
that the magnitude of $K_{u001}$ is negligibly affected by
$\bm{\epsilon}'$, the results should already match the
experimental data, and indeed they do. Although the spread of the
experimental points in Fig.~\ref{fig:All_K}a is significantly
larger than that of the theoretical curves, the correspondence
between the computed and experimental values is good, and it has
been achieved without introducing into the model any adjustable
parameters. This has been only possible by including  the hole
liquid magnetization in the calculation of $x_{\text{eff}}$. When
$M_h$ is disregarded, the experimental values of $K_{u001}$ are
systematically above the \emph{maxima} of the theoretical curves.

As already mentioned, in the case of the (001) (Ga,Mn)As films an
additional low-symmetry term has to be introduced into the
Hamiltonian in order to reproduce the experimentally observed
uniaxial in-plane magnetic anisotropy. In a general case of an
arbitrarily oriented substrate, there are three anisotropy
constants of this kind, $K_{xy}$, $K_{yz}$, and $K_{xz}$. In the case of
(113) (Ga,Mn)As, the symmetry requires that $K_{xz} = K_{yz}$, the
assumption confirmed by the description of the experimental
results. The two relevant anisotropy constants $K_{xy}$ and
$K_{xz}$ are presented in Figs.~\ref{fig:All_K}(b) and~(c), and
are seen to be non-zero even in the absence of a symmetry lowering
perturbation, $\epsilon'_{xy} = 0$. The computed
magnitude of $K_{xy}$ for $\epsilon'_{xy} = 0$ yields an
acceptable agreement with the experimental data. We note, moreover,
that an exact match is possible when allowing for non-zero values
of $\epsilon'_{xy} \cong 0.001$\% for sample S1 and 0.01\% for
samples S2 and S3.

In contrast, according to Fig.~\ref{fig:All_K}(c), the theoretical
description of the experimental values of the $K_{xz}$ anisotropy
constant requires a quite sizable value of the corresponding strain component
$\epsilon'_{xz} = -0.1\%$. Thus, remarkably and contrary to the case of
(001) (Ga,Mn)As, one barely needs any in-plane shear strain to
reproduce in-plane uniaxial anisotropy, whereas for the
out-of-plane component two times stronger shear strain is needed
comparing  to the (001) case. This finding should be taken as a
strong evidence that the in-growth surface reconstruction, and a
related with it orientational preferences of Mn incorporation must
play a decisive role in the mechanism leading to the lowering of
magnetic symmetry.

Finally, we turn to the case of the cubic anisotropy constant
$K_C$, shown in Fig.~\ref{fig:All_K}d. Since $K_C$ shows only a
small sensitivity to $\bm{\epsilon}'$, we present the results of
computations only for $\bm{\epsilon}' = 0$. We find that similarly
to the (001) case,\cite{Sawicki:2004_PRB} the present theory
underestimates the magnitude of $K_C$, particularly in the low
hole concentration region, where the theoretically expected change
of sign of $K_C$ is not observed experimentally.  The origin of
this discrepancy, and in particular its relation to the symmetry
lowering perturbation is presently unknown.

We have also examined theoretically how the particular anisotropy
constants depend on magnetization $M$. As could be expected, and
in a qualitative agreement with the experimental finding shown in
Fig.~\ref{fig:magnetic anisotropyconstants}, the uniaxial
anisotropy constants $K_{u001}$, $K_{xy}$, and $K_{xz}$ (or
equivalently $K_{u1}$ and $K_{u\overline110}$) are proportional to
$M^2$, whereas $K_C$ to $M^4$, except to the hole concentration
region in the immediate vicinity of the sign change.

\section{Conclusions}

We have investigated the magnetic properties of as-grown and
annealed (Ga,Mn)As layers grown by MBE on GaAs substrates of the
$(113)$ orientation and provided the most complete to date
description of the magnetic anisotropy in the whole temperature
range up to $T_{\mathrm{C}}$. At higher temperatures the
$[\overline110]$ direction is the easy magnetization axis before
and after annealing. At low temperature the spin reorientation
transition to a pair of easy axes near the $[100]$ and $[010]$
directions takes place and to a first approximation, magnetization
behavior as a function of temperature is similar to that observed
in $(001)$ (Ga,Mn)As in the absence of an external magnetic
field.\cite{Wang:2005_PRL} However, the magnetization vector
resides in a plane close to (001) plane only for a low value of
the hole concentration. When it increases, the plane rotates along
$[\overline110]$ towards the sample face (113), and the two cubic easy
directions move towards the $[33\overline2]$ direction.

We have estimated the values of magnetic anisotropy constants by
fitting our phenomenological model to the hysteresis loops
measured by SQUID. The comparison to results of FMR measurements
confirms the correctness of this approach. The obtained values of
the cubic and uniaxial in-plane magnetic anisotropy constants are
proportional to $M^4$ and $M^2$, respectively.  Inflections from
the $M^2$ dependence of the out-of-plane uniaxial anisotropy
constant indicate a proximity to another spin reorientation
transition at which the out-of-plane axis becomes easy on lowering
temperature. It has been evidenced by MOKE that it is possible to
reverse the out-of-plane magnetization component by applying an
in-plane magnetic field.

For the hole and effective Mn concentrations determined from the
values of saturation magnetization and the total Mn concentration,
the $p-d$ Zener model explains, with no adjustable parameters, the
magnitude of the Curie temperature as well as the sign and
magnitude of the uniaxial $[001]$ anisotropy constant $K_{u001}$
caused by biaxial strain. At the same time, however, the predicted
values of the cubic anisotropy constant are smaller than those
found experimentally in the hole concentration range studied here.
For the substrate orientation in question there are two additional
non-zero second order ($l = 2$) components $K_{xy}$ and $K_{xz}$.
The comparison of their experimental and theoretical values points
to the presence of an additional shear strain. The non-vanishing
components of this additional strain are  $\epsilon_{xz}' =
\epsilon_{yz}' \approx -0.1$\%, in contrast with $(001)$ samples,
for which a non-zero value of $\epsilon_{xy}'\approx = 0.05\%$ has
to be assumed in order to explain the experimental data. This
finding provides a hint that a preferential Mn incorporation
during the growth process accounts for the mysterious lowering of
the (Ga,Mn)As symmetry.

\section*{Acknowledgments}

The work was supported by EU FunDMS Advanced Grant of the European
Research Council within the "Ideas" 7th Framework Programme,
InTechFun (POIG.01.03.01-00-159/08), SemiSpinNet
(PITN-GA-2008-215368) and Polish MNiSW 2048/B/H03/2008/34 grant.
We thank M.~Kisielewski and M.~Maziewski for valuable discussions
on optical measurements.

\bibliography{vstef311_new_style,Baza_nowa_01.02.2010}

\end{document}